\documentclass[final,1p,times,twocolumn]{elsarticle}
\usepackage{lineno,hyperref,amsmath,amsthm,amssymb,amsfonts,ragged2e,color,subfig}
\journal{``Contribution to Plasma Physics''}
\begin{document}
\begin{frontmatter}
\title{Modulational instability of dust-ion-acoustic waves in pair-ion plasma having non-thermal non-extensive electrons}
\author{M.K. Islam$^{*,1}$, A.A. Noman$^{**,1}$, J. Akter$^{***,1}$, N.A. Chowdhury$^{\dag,2}$, A. Mannan$^{\dag\dag,1,3}$,\\
T.S. Roy$^{\ddag,4}$, M. Salahuddin$^{\S,1}$, and A.A. Mamun$^{\S\S,1}$}
\address{$^1$ Department of Physics, Jahangirnagar University, Savar, Dhaka-1342, Bangladesh\\
$^2$ Plasma Physics Division, Atomic Energy Centre, Dhaka-1000, Bangladesh\\
$^3$ Institut f\"{u}r Mathematik, Martin Luther Universit\"{a}t Halle-Wittenberg, Halle, Germany\\
$^4$ Department of Physics, Bangladesh University of Textiles, Tejgaon Industrial Area, Dhaka, Bangladesh\\
e-mail: $^*$islam.stu2018@juniv.edu, $^{**}$noman179physics@gmail.com, $^{***}$akter277phy@gmail.com,\\
$^{\dag}$nurealam1743phy@gmail.com, $^{\dag\dag}$abdulmannan@juniv.edu, $^{\ddag}$tanu.jabi@gmail.com,\\
 $^\S$su\_2960@juniv.edu, $^{\S\S}$mamun\_phys@juniv.edu}
\begin{abstract}
The modulational instability (MI) criteria of dust-ion-acoustic (DIA) waves (DIAWs)
have been investigated in a four-component pair-ion plasma having inertial pair-ions, inertialess non-thermal non-extensive
electrons, and immobile negatively charged massive dust grains. A nonlinear Schr\"{o}dinger equation (NLSE) is derived
by using reductive perturbation method. The nonlinear and dispersive coefficients of the NLSE
can predict the modulationally stable and unstable parametric regimes of DIAWs and associated first and
second order DIA rogue waves (DIARWs). The MI growth rate and the configuration of the DIARWs
are examined, and it is found that the MI growth rate increases (decreases) with
increasing the number density of the negatively charged dust grains in the presence (absence) of the negative ions. It is also observed that
the amplitude and width of the DIARWs increase (decrease) with the negative (positive) ion mass.
The implications of the results to laboratory and space plasmas are briefly discussed.
\end{abstract}
\end{frontmatter}
\section{Introduction}
\label{1sec:introduction}
The ubiquitous existence of pair-ion (PI) in astrophysical environments such as
upper regions of Titan's atmosphere \cite{Massey1976,Sabry2009,Abdelwahed2016,Misra2009,Mushtaq2012}, cometary comae \cite{Chaizy1991},
the ($H^+$, $O_2^-$) and ($H^+$, $H^-$) plasmas in the D and F-regions of Earth's
ionosphere \cite{Massey1976,Sabry2009,Abdelwahed2016,Misra2009,Mushtaq2012}, and also
in the laboratory experiments namely, neutral beam sources \cite{Bacal1979}, plasma processing reactors \cite{Gottscho1986},
and Fullerene ($C^+$, $C^-$) \cite{Oohara2003} have enormously attracted the plasma physicists
to examine nonlinear electrostatic modes in PI plasma. The properties of the laboratory PI plasma
have also been observed by a number of authors \cite{Oohara2003,Hatakeyama2005,Oohara2005}.
The dynamics of the PI plasma system and the configuration of the associated nonlinear electrostatic
structures have been rigorously changed by the presence of negatively charged massive dust grains.
Misra \cite{Misra2009} examined dust-ion-acoustic (DIA) shock waves in PI plasma in the presence of immobile negatively charged
dust grains, and reported that the height of the shock front seems to decrease with the number density of dust grains.
Mushtaq \textit{et al.} \cite{Mushtaq2012} considered a plasma model having inertialess non-thermal electrons,
inertial positive and negative ions, and immobile negatively charged massive dust grains to investigate
DIA solitary waves, and found that the amplitude of the solitary waves increases with
increasing the number density and charge state of the negative dust grains.

The highly energetic particles associated with long tail in space plasmas have been
identified by Viking spacecraft \cite{Bostrom1992} and Freja satellite \cite{Dovner1994}.
Cairns \textit{et al.} \cite{Cairns1995} proposed non-thermal velocity distribution for
these energetic particles in space plasmas, and observed electrostatic solitary structures. On the other hand,
Tsallis \cite{Tsallis1988} introduced non-extensive $q$-distribution for explaining the high energy tail in
space plasma. The parameter $q$ in the non-extensive $q$-distribution describes the deviation of the plasma particles from the
thermally equilibrium state. A number of authors \cite{Tribeche2012,Williams2013,Dutta2017} have considered
non-thermal non-extensive distribution for investigating nonlinear electrostatic waves. Dutta and Sahu \cite{Dutta2017}
examined DIA waves (DIAWs) by considering inertialess non-thermal non-extensive electrons, inertial PI, and stationary negatively
charged dust grains, and highlighted that the angular frequency of the DIAWs
decreases with  non-extensivity of electrons but increases with non-thermality of electrons.

Rogue waves (RWs) are the results of modulational instability (MI) of the carrier waves, and are governed by the rational solution of standard
nonlinear Schr\"{o}dinger equation (NLSE) \cite{Akhtar2019,Sun2018,Shalini2017}, and have been observed in fiber optics \cite{Kibler2010}, water waves \cite{Chabchoub2011}, and plasmas \cite{Bailung2011}, etc. Bailung \textit{et al.} \cite{Bailung2011} experimentally observed the RWs in a multi-component plasma, and found that the amplitude of the RWs is three times greater in comparison with surrounding normal waves.
Abdelwahed \textit{et al.} \cite{Abdelwahed2016} observed the effects of the negative ion on the RWs
in PI plasma and determined that the negative ion number density increases the nonlinearity as well as the amplitude and
width of the RWs in PI plasma. Bains \textit{et al.} \cite{Bains2011} analyzed the MI of ion-acoustic waves (IAWs)
and identified that the critical wave number ($k_c$) increases with $q$. El-Labany \textit{et al.} \cite{El-Labany2012}
investigated the MI of the IAWs and associated RWs in a three-component plasma system having inertialess
iso-thermal electrons and inertial PI, and found that the amplitude and width of the RWs increase with the number density and mass of the negative ion.
Javidan and Pakzad \cite{Javidan2014} considered a three-component dusty plasma containing inertial ions, inertialess electrons, and immobile dust grains,
and examined the instability of the IAWs, and demonstrated that the instability growth rate decreases with the increase in the value of negative dust number density.
To the best knowledge of the authors, no attempt
has been made to study the MI of DIAWs and associated DIA RWs (DIARWs) in a four-component plasma having inertialess non-thermal non-extensive electrons,
inertial positive and negative ions, and stationary negatively charged massive
dust grains. The aim of the present investigation is, therefore, to derive NLSE and investigate MI criteria of DIAWs in a four-component plasma,
and to observe the effects of various plasma parameters to the formation of first and second order DIARWs.

This manuscript is organized in the following way: The governing equations are
described in Sec. \ref{1sec:Governing Equations}. The standard NLSE is derived
in Sec. \ref{1sec:Derivation of NLSE}. The MI of the DIAWs and associated DIARWs
are examined in Sec. \ref{1sec:Modulational Instability and Rouge Waves}. The
conclusion of our present work is provided in Sec. \ref{1sec:Conclusion}.
\section{Governing Equations}
\label{1sec:Governing Equations}
We consider a four-component PI plasma medium having inertial positive ion (charge $q_+=eZ_+$
and mass $m_+$), inertial negative ion (charge $q_-=-eZ_-$ and mass $m_-$), inertialess non-thermal
non-extensive electron (charge $q_e=-e$ and mass $m_e$), and immobile negatively charged massive dust grains (charge $q_d=-eZ_d$ and mass $m_d$).
Overall, the charge neutrality condition for our plasma model can be written as $Z_+n_{+0}=n_{e0}+Z_-n_{-0}+ Z_dn_{d0}$,
where $Z_+$ ($Z_-$) is the charge state of the positive (negative) ion, and $Z_d$ is the negatively charged massive dust grains' charge state.
The propagation of the DIAW is governed by the following equations:
\begin{eqnarray}
&&\hspace*{-1.3cm}\frac{\partial n_{+}}{\partial t}+\frac{\partial}{\partial x}(n_{+}u_{+})=0,
\label{1eq:1}\\
&&\hspace*{-1.3cm}\frac{\partial u_{+}}{\partial t}+u_{+}\frac{\partial u_{+}}{\partial x}=-\frac{\partial\phi}{\partial x},
\label{1eq:2}\\
&&\hspace*{-1.3cm}\frac{\partial n_{-}}{\partial t}+\frac{\partial}{\partial x}(n_{-}u_{-})=0,
\label{1eq:3}\\
&&\hspace*{-1.3cm}\frac{\partial u_{-}}{\partial t}+u_{-}\frac{\partial u_{-}}{\partial x}=\mu_1\frac{\partial\phi}{\partial x},
\label{1eq:4}\\
&&\hspace*{-1.3cm}\frac{\partial^2\phi}{\partial x^2}=(1-\mu_2-\mu_3)n_e-n_{+}+\mu_2n_{-}+\mu_3,
\label{1eq:5}\
\end{eqnarray}
where $n_+$ $(n_-)$ is the positive (negative) ion number density normalized by it's equilibrium
value $n_{+0} $ $(n_{-0})$; $u_+$ $(u_-)$ is the positive (negative) ion fluid speed normalized by
the ion-acoustic wave speed $C_+=(Z_+ k_BT_e/m_+)^{1/2}$ with $T_e$ being the non-thermal non-extensive
electron temperature and $k_B$ being the Boltzmann constant; $\phi$ is the electrostatic wave potential normalized
by $k_BT_e/e$; the time and space variables are normalized by ${\omega^{-1}_{p+}}=(m_+/4\pi {Z_+}^2 e^2 n_{+0})^{1/2}$
and $\lambda_{D+}=(k_BT_e/4 \pi Z_+ e^2 n_{+0})^{1/2}$, respectively. Other plasma parameters can be written as $\mu_1=Z_-m_+/Z_+m_-$, $\mu_2=Z_-n_{-0}/Z_+n_{+0}$, and $\mu_3=Z_d n_{d0}/Z_+ n_{+0}$. Now, the expression for the number density of electrons following non-thermal non-extensive distribution \cite{Tribeche2012} can be written as
\begin{eqnarray}
&&\hspace{-0.8cm}n_e=\big[1+A\phi+B\phi^2\big]\times\big[1+(q-1)\phi\big]^{\frac{(q+1)}{2(q-1)}},
\label{1eq:6}\
\end{eqnarray}
where the parameter $q$ stands for the strength of non-extensive system and the
coefficients $A$  and $B$ are defined by $A=-16q\alpha/(3-14q+15q^2+12\alpha)$
and $B=-A(2q-1)$. Here $\alpha$ is a parameter determining the number of
non-thermal electrons in the model. Williams \textit{et al.} \cite{Williams2013} discussed the
range and the validity of ($q$, $\alpha$) for solitons. In the limiting case $(q\rightarrow1$ and $\alpha= 0$),
the above distribution reduces to the well-known Maxwell-Boltzmann velocity distribution.
For ($q\rightarrow1$ and  $\alpha\neq 0$), the above distribution reduces to Cairns' distribution.
Now, by substituting Eq. \eqref{1eq:6} into Eq. \eqref{1eq:5} and expanding up to
third order of $\phi$, we get
\begin{eqnarray}
&&\hspace*{-1.3cm}\frac{\partial^2\phi}{\partial x^2}+n_{+}=1+\mu_2+\mu_2n_-+\gamma_1\phi+\gamma_2\phi^2+\gamma_3\phi^3+\cdot\cdot\cdot,
\label{1eq:7}\
\end{eqnarray}
where
\begin{eqnarray}
&&\hspace*{-1.3cm}\gamma_1=[(1-\mu_2-\mu_3)(2A+q+1)]/2,~~~~~~~\gamma_2=[(1-\mu_2-\mu_3)\{8B+(q+1)(4A-q+3)\}]/8,
\nonumber\\
&&\hspace*{-1.3cm}\gamma_3=[(1-\mu_2-\mu_3)\{24B(q+1)-(q+1)(q-3)(6A-3q+5)\}]/48.
\nonumber\
\end{eqnarray}
The terms containing $\gamma_1$, $\gamma_2$, and $\gamma_3$ in Eq. \eqref{1eq:7}
are due to the contribution of the non-thermal non-extensive electrons.
\section{Derivation of the NLSE}
\label{1sec:Derivation of NLSE}
\begin{figure}[t!]
\centering
\includegraphics[width=130mm]{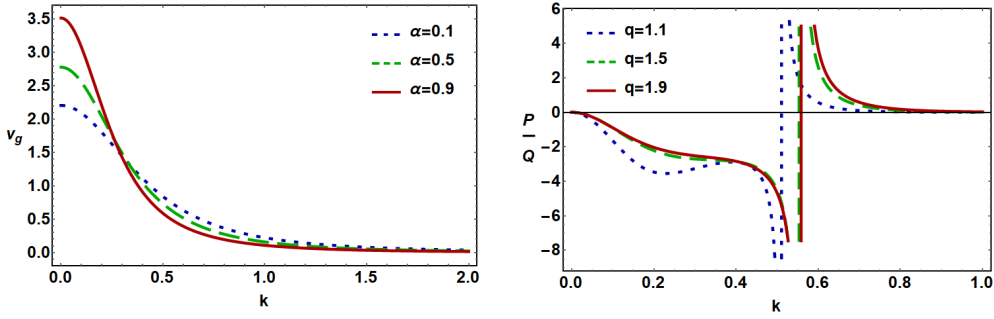}
\caption{The variation of $v_g$ with $k$ for different values of $\alpha$ when $q=1.5$ (left panel),
and the variation of $P/Q$ with $k$ for different values of $q$ when $\alpha=0.5$ (right panel).
Other plasma parameters are $\mu_1=0.5$, $\mu_2=0.7$, and $\mu_3=0.05$.}
\label{1Fig:2}
\end{figure}

To study the MI of the DIAWs, we want to derive the NLSE by employing the reductive perturbation method (RPM)
and for that case, first we can write the stretched co-ordinates in the form \cite{Khondaker2019,Noman2019,Shikha2021}
\begin{eqnarray}
&&\hspace*{-1.3cm}\xi=\epsilon(x-v_gt),
\label{1eq:8}\\
&&\hspace*{-1.3cm}\tau=\epsilon^2t,
\label{1eq:9}\
\end{eqnarray}
where $v_g$ is the group speed and $\epsilon$ ($0<\epsilon<1$) is a small parameter measuring
the weakness of the dispersion. Then, we can write the dependent variables as \cite{Khondaker2019,Noman2019,Shikha2021}
\begin{eqnarray}
&&\hspace*{-1.3cm} n_+=1+\sum_{m=1}^{\infty} \epsilon^{m} \sum_{l=-\infty}^{\infty} n_{+l}^{(m)} (\xi,\tau) \exp[il(kx-\omega t)],
\label{1eq:10}\\
&&\hspace*{-1.3cm} n_-=1+\sum_{m=1}^{\infty} \epsilon^{m} \sum_{l=-\infty}^{\infty} n_{-l}^{(m)} (\xi,\tau) \exp[il(kx-\omega t)],
\label{1eq:11}\\
&&\hspace*{-1.3cm} u_+=\sum_{m=1}^{\infty} \epsilon^{m} \sum_{l=-\infty}^{\infty} u_{+l}^{(m)} (\xi,\tau) \exp[il(kx-\omega t)],
\label{1eq:12}\\
&&\hspace*{-1.3cm} u_-=\sum_{m=1}^{\infty} \epsilon^{m} \sum_{l=-\infty}^{\infty} u_{-l}^{(m)} (\xi,\tau) \exp[il(kx-\omega t)],
\label{1eq:13}\\
&&\hspace*{-1.3cm} \phi=\sum_{m=1}^{\infty} \epsilon^{m} \sum_{l=-\infty}^{\infty} \phi_{l}^{(m)} (\xi,\tau) \exp[il(kx-\omega t)],
\label{1eq:14}\
\end{eqnarray}
where $k$ and $\omega$ are the real variables representing the carrier wave number and frequency,
respectively. The derivative operators can be written as
\begin{eqnarray}
&&\hspace*{-1.3cm}\frac{\partial}{\partial t}\rightarrow\frac{\partial}{\partial t}-\epsilon v_g \frac{\partial}{\partial \xi}+ \epsilon^2\frac{\partial}
{\partial \tau},
\label{1eq:15}\\
&&\hspace*{-1.3cm}\frac{\partial}{\partial x}\rightarrow\frac{\partial}{\partial x}+\epsilon\frac{\partial}{\partial \xi}.
\label{1eq:16}\
\end{eqnarray}
Now, by substituting Eqs. \eqref{1eq:8}-\eqref{1eq:16} into Eqs. \eqref{1eq:1}-\eqref{1eq:4} and \eqref{1eq:7}, and
collecting the terms containing $\epsilon$, the first order (when $m=1$ with $l=1$) reduced equations can be written as
\begin{eqnarray}
&&\hspace*{-1.3cm}u_{+1}^{(1)} =\frac{k}{\omega}\phi_1^{(1)},~~~~~n_{+1}^{(1)} = \frac{k^2}{\omega^2}\phi_1^{(1)},~~~~~u_{-1}^{(1)} =-\frac{\mu_1k}{\omega}\phi_1^{(1)},~~~~~n_{-1}^{(1)} = -\frac{\mu_1k^2}{\omega^2}\phi_1^{(1)},
\nonumber\
\end{eqnarray}
these relations provide the dispersion relation of DIAWs
\begin{eqnarray}
&&\hspace*{-1.3cm}\omega^2=\frac{k^2\big(1+\mu_1\mu_2\big)}{k^2+\gamma_1}.
\label{1eq:17}\
\end{eqnarray}
\begin{figure}[t!]
\centering
\includegraphics[width=130mm]{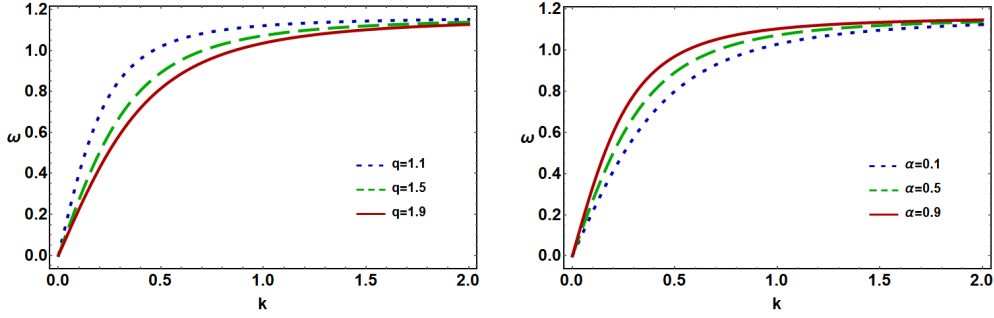}
\caption{The variation of $\omega$ with $k$ for different values of $q$ when $\alpha=0.5$ (left panel), and the variation of $\omega$
with $k$ for different values of $\alpha$ when $q=1.5$ (right panel). Other plasma parameters are  $\mu_1=0.5$, $\mu_2=0.7$, and
$\mu_3=0.05$.}
\label{1Fig:1}
\end{figure}
We have numerically analyzed Eq. \eqref{1eq:17} to examine the dispersion properties of DIAWs for
different values of $q$ and $\alpha$. The results are displayed in Fig. \ref{1Fig:1}, which shows that (a) for small
wave number, the angular frequency of the DIAWs exponentially increases and for large wave number, the dispersion
curves become saturated (both left and right panel); (b) the $\omega$ decreases with $q$ (left panel), and this result
is a good agreement with the result of Dutta and Sahu \cite{Dutta2017}; and (c) as we increase the non-thermality of
the electron then the $\omega$ also increases (right panel), and this result also coincides with the work of Dutta
and Sahu \cite{Dutta2017}. The second order equations (when $m=2$ with $l=1$) are given by
\begin{eqnarray}
&&\hspace*{-1.3cm}u_{+1}^{(2)} = \frac{k}{\omega}\phi_1^{(2)}+\frac{i(v_gk-\omega)}{\omega^2}\frac{\partial \phi_1^{(1)}}{\partial \xi},~~~~~~~~~~~~~~~n_{+1}^{(2)} = \frac{k^2}{\omega^2}\phi_1^{(2)}+\frac{2ik(v_gk-\omega)}{\omega^3}\frac{\partial \phi_1^{(1)}}{\partial \xi},
\nonumber\\
&&\hspace*{-1.3cm}u_{-1}^{(2)} = -\frac{k\mu_1}{\omega}\phi_1^{(2)}-\frac{i\mu_1(v_gk-\omega)}{\omega^2}\frac{\partial \phi_1^{(1)}}{\partial \xi},~~~~~n_{-1}^{(2)} = -\frac{\mu_1k^2}{\omega^2}\phi_1^{(2)}-\frac{2ik\mu_1(v_gk-\omega)}{\omega^3}\frac{\partial \phi_1^{(1)}}{\partial \xi},
\nonumber\
\end{eqnarray}
and with the compatibility condition, the group velocity of the DIAWs can be written as
\begin{eqnarray}
&&\hspace*{-1.3cm}v_g=\frac{\omega(1-\omega^2+\mu_1\mu_2)}{k(1+\mu_1\mu_2)}.
\label{1eq:18}\
\end{eqnarray}
The variation of $v_g$ with $k$ can be seen from the left panel of Fig. \ref{1Fig:2},
and it is clear from this figure that the group velocity increases with increasing the value of $\alpha$.
Therefore, one can conclude that the group velocity of DIAWs increases as the non-thermality of electrons increases.
The coefficients of $\epsilon$ when  $m=2$ with $l=2$ provides the second
order harmonic amplitudes which are found to be proportional
to $|\phi_1^{(1)}|^2$
\begin{eqnarray}
&&\hspace*{-1.3cm}n_{+2}^{(2)}=\mu_4 |\phi_1^{(1)}|^2,~~~~~u_{+2}^{(2)}=\mu_5 |\phi_1^{(1)}|^2,~~~~~n_{-2}^{(2)}=\mu_6 |\phi_1^{(1)}|^2,~~~~~u_{-2}^{(2)}=\mu_7 |\phi_1^{(1)}|^2,~~~~~\phi_{2}^{(2)}=\mu_8 |\phi_1^{(1)}|^2,
\label{1eq:19}\
\end{eqnarray}
where
\begin{eqnarray}
&&\hspace*{-0.8cm}\mu_4=\frac{3k^4+2k^2\omega^2\mu_8}{2\omega^4},~~~~~~~\mu_5=\frac{k^3+2k\omega^2\mu_8}{2\omega^3},
~~~~~~~\mu_6=\frac{3k^4\mu_1^2-2k^2\omega^2\mu_1\mu_8}{2\omega^4},
\nonumber\\
&&\hspace*{-0.8cm}\mu_7=\frac{k^3\mu_1^2-2k\omega^2\mu_1\mu_8}{2\omega^3},~~~~~~~\mu_8=\frac{3k^4-2\gamma_2\omega^4-3\mu_1^2\mu_2k^4}{2\omega^2(4k^2\omega^2+\gamma_1\omega^2-k^2-k^2\mu_1\mu_2)}.
\nonumber\
\end{eqnarray}
Now, $m=3$ with $l=0$ and $m=2$ with $l=0$ lead to zeroth harmonic modes as follows:
\begin{eqnarray}
&&\hspace*{-1.3cm}n_{+0}^{(2)}=\mu_9 |\phi_1^{(1)}|^2,~~~~~u_{+0}^{(2)}=\mu_{10} |\phi_1^{(1)}|^2,~~~~n_{-0}^{(2)}=\mu_{11} |\phi_1^{(1)}|^2,~~~u_{-0}^{(2)}=\mu_{12} |\phi_1^{(1)}|^2,~~~~\phi_{0}^{(2)}=\mu_{13} |\phi_1^{(1)}|^2,
\label{1eq:20}\
\end{eqnarray}
where
\begin{eqnarray}
&&\hspace*{-0.8cm}\mu_9=\frac{2k^3v_g+k^2\omega+\mu_{13}\omega^3}{v_g^2\omega^3},~~~~~~~\mu_{10}=\frac{k^2+\mu_{13}\omega^2}{v_g\omega^2},~~~~~~~\mu_{11}=\frac{2k^3v_g\mu_1^2+k^2\mu_1^2\omega-\mu_1\mu_{13}\omega^3}{v_g^2\omega^3},
\nonumber\\
&&\hspace*{-0.8cm}\mu_{12}=\frac{k^2\mu_1^2-\mu_1\mu_{13}\omega^2}{v_g\omega^2},~~~~~~~\mu_{13}=\frac{2\gamma_2v_g^2\omega^3+k^2\mu_1^2\mu_2(2kv_g+\omega)-k^2(2kv_g+\omega)}{\omega^3(1+\mu_1\mu_2-\gamma_1v_g^2)}.
\nonumber\
\end{eqnarray}
Finally, the third harmonic modes, when $m=3$ and $l=1$, with the help of Eqs. \eqref{1eq:17}-\eqref{1eq:20} give a
set of equations which can be reduced to the standard NLSE:
\begin{eqnarray}
&&\hspace*{-1.3cm}i \frac{\partial \Phi}{\partial \tau}+P\frac{\partial^2 \Phi}{\partial \xi^2}+Q|\Phi|^2\Phi=0,
\label{1eq:21}\
\end{eqnarray}
where $\Phi=\phi_1^{(1)}$, for simplicity. $P$ and $Q$ are the dispersion and nonlinear coefficients of the NLSE, respectively,
\begin{eqnarray}
&&P=\frac{3v_g(v_gk-\omega)}{2\omega k},
\label{1eq:22}\\
&&Q=\frac{2\gamma_2\omega^3(\mu_8+\mu_{13})-k^2\omega(\mu_4+\mu_9+\mu_1\mu_6+\mu_1\mu_{11})+3\gamma_3\omega^3-2k^3(\mu_5+\mu_{10}+\mu_1\mu_7+\mu_1\mu_{12})}{2k^2(1+\mu_1\mu_2)}.
\label{1eq:23}\
\end{eqnarray}
The space and time evolution of the DIAWs in PI plasma are directly governed by the coefficients $P$ and $Q$, and indirectly governed by
the different plasma parameters such as $q$, $\alpha$, $\mu_1$, $\mu_2$, and $\mu_3$.
\begin{figure}[t!]
\centering
\includegraphics[width=130mm]{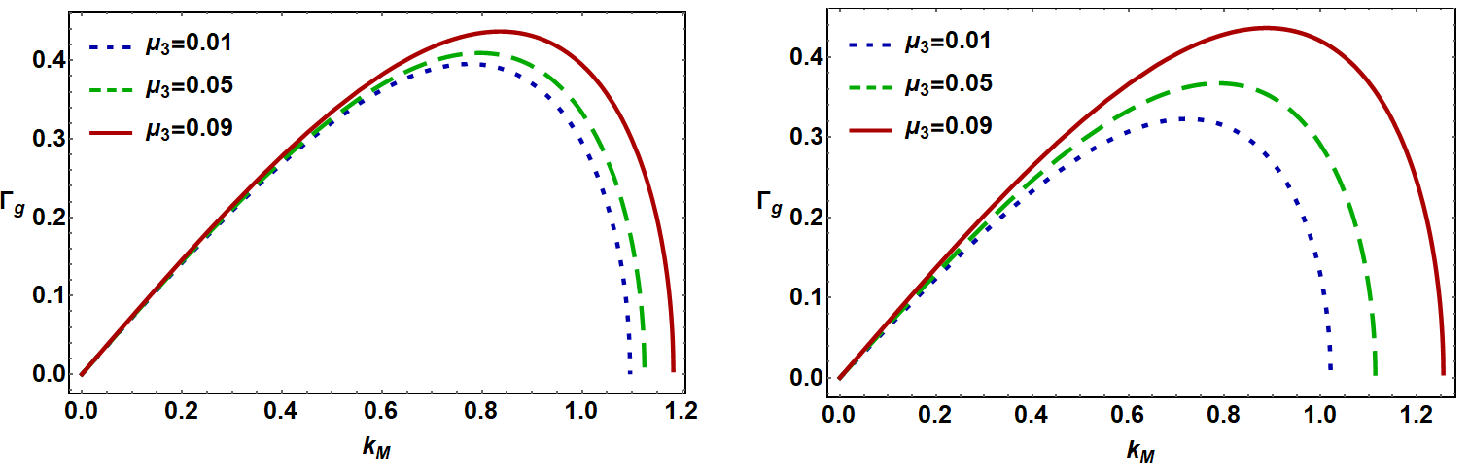}
\caption{The variation of $\Gamma_g$ with $k_M$ for different values of  $\mu_3$ when $q=1$ and $\alpha=0$ (left panel),
and the variation of $\Gamma_g$ with $k_M$ for different values of $\mu_3$ when $q=1.5$ and $\alpha=0.5$ (right panel).
Other plasma parameters are $k=0.7$, $\phi_0=0.5$, $\mu_1=0.5$, and $\mu_2=0.7$.}
\label{1Fig:3}
\end{figure}
\begin{figure}[t!]
\centering
\includegraphics[width=130mm]{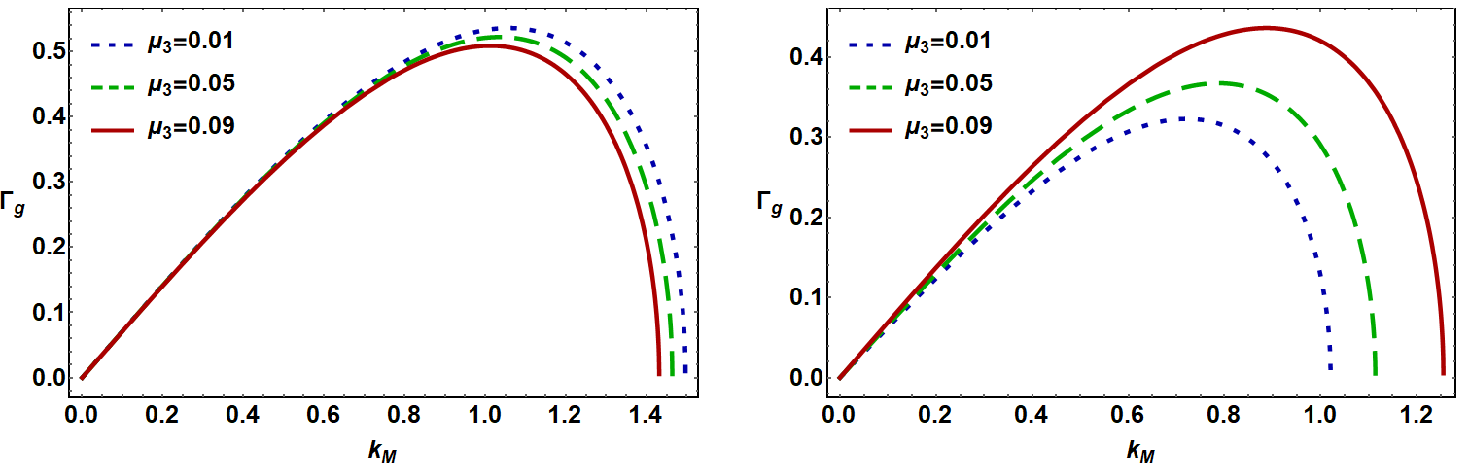}
\caption{The variation of $\Gamma_g$ with $k_M$ for different values of  $\mu_3$ when $\mu_2=0$ (left panel),
and the variation of $\Gamma_g$ with $k_M$ for different values of $\mu_3$ when $\mu_2=0.5$ (right panel).
Other plasma parameters are $k=0.7$, $\phi_0=0.5$, $\alpha=0.5$, $q=1.5$, and $\mu_1=0.5$.}
\label{1Fig:4}
\end{figure}
\begin{figure}[t!]
\centering
\includegraphics[width=130mm]{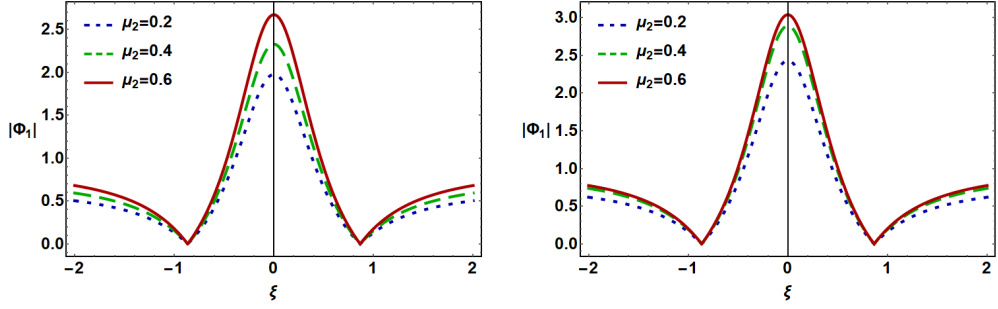}
\caption{The variation of $|\Phi_1|$ with $\xi$ for different values of  $\mu_2$ when $q=1$ and $\alpha=0$ (left panel),
and the variation of $|\Phi_1|$ with $\xi$ for different values of $\mu_2$ when $q=1.5$ and $\alpha=0.5$ (right panel).
Other plasma parameters are $k=0.7$, $\tau=0$, $\mu_1=0.5$, and $\mu_3=0.05$.}
\label{1Fig:5}
\end{figure}
\begin{figure}[t!]
\centering
\includegraphics[width=130mm]{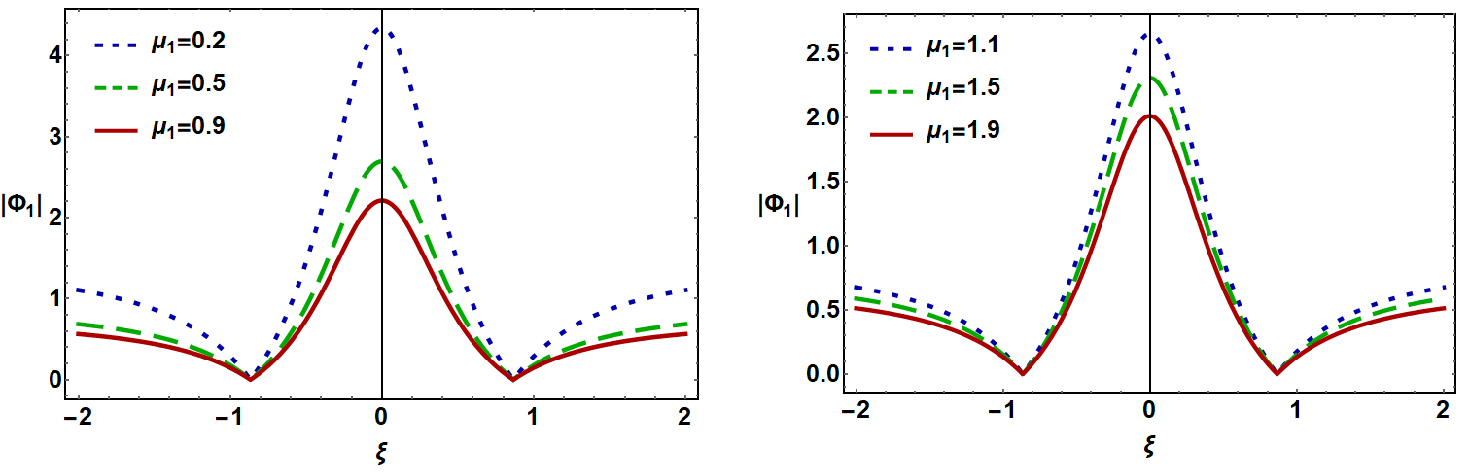}
\caption{The variation of $|\Phi_1|$ with $\xi$ for different values of $\mu_1$ (both left and right panel).
Other plasma parameters are $k=0.7$, $\tau=0$, $\alpha=0.5$, $q=1.5$, $\mu_2=0.7$, and $\mu_3=0.05$.}
\label{1Fig:6}
\end{figure}
\begin{figure}[t!]
\centering
\includegraphics[width=130mm]{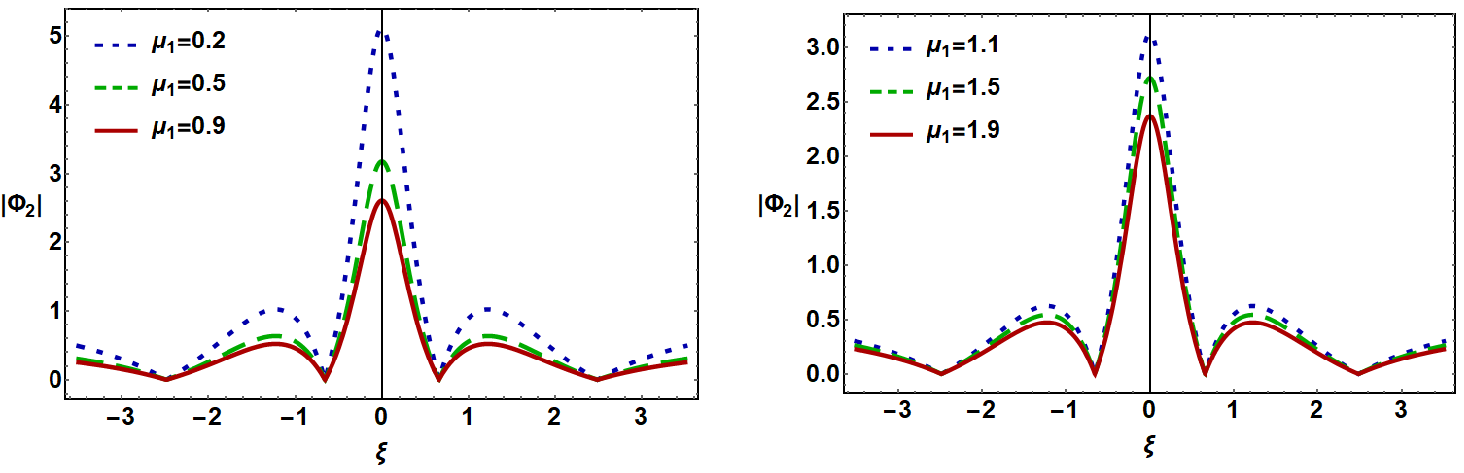}
\caption{The variation of $|\Phi_2|$ with $\xi$ for different values of $\mu_1$ (both left and right panel).
Other plasma parameters are $k=0.7$, $\tau=0$, $\alpha=0.5$, $q=1.5$, $\mu_2=0.7$, and $\mu_3=0.05$.}
\label{1Fig:7}
\end{figure}
\begin{figure}[t!]
\centering
\includegraphics[width=80mm]{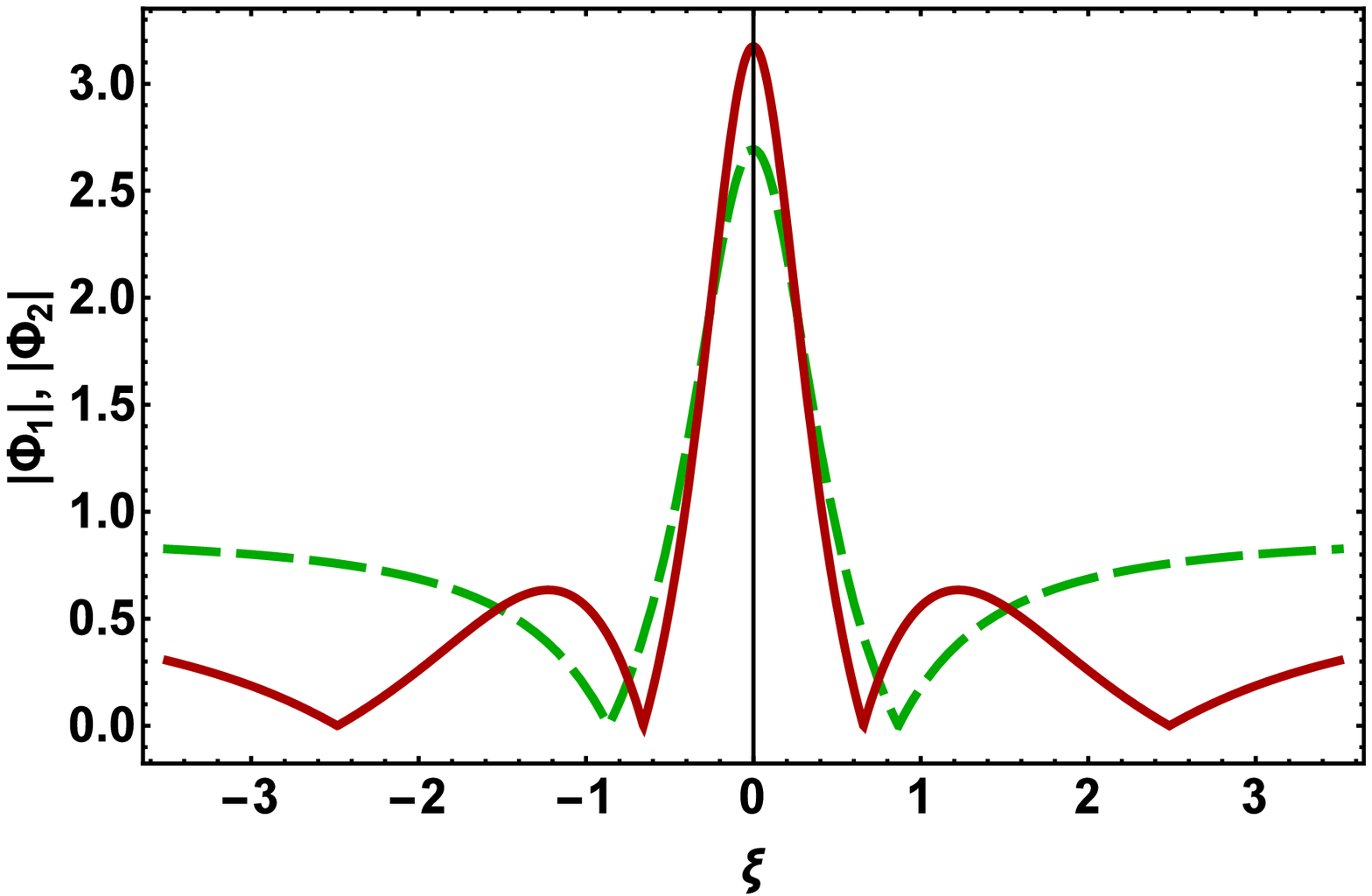}
\caption{The variation of first-order (dashed green curve) and
second-order (solid red curve) rational solutions of NLSE at
$k=0.7$, $\tau=0$, $\alpha=0.5$, $q=1.5$, $\mu_1=0.5$, $\mu_2=0.7$, and $\mu_3=0.05$.}
\label{1Fig:8}
\end{figure}
\section{Modulational Instability and Rogue Waves}
\label{1sec:Modulational Instability and Rouge Waves}
\label{2sec:Modulational instability and rogue waves}
The stable and unstable parametric regimes of the DIAWs are determined by the sign of the dispersion ($P$)
and nonlinear ($Q$) coefficients of the standard NLSE \cite{Kourakis2003,Kourakis2005,Fedele2002,Sultana2011}.
When $P$ and $Q$ have same sign (i.e., $P/Q>0$),
the evolution of the DIAWs amplitude is modulationally unstable. On the other hand, when $P$ and $Q$ have
opposite sign (i.e., $P/Q<0$), the DIAWs are modulationally stable in the presence of external perturbations.
The plot of $P/Q$ against $k$ yields stable and unstable parametric regimes of DIAWs.
The point, at which transition of $P/Q$ curve intersects with $k$-axis, is known as threshold
or critical wave number $k$ ($=k_c$) \cite{Kourakis2003,Kourakis2005,Fedele2002,Sultana2011}.
We have numerically analyzed the variation of $P/Q$ with $k$ for different
values of $q$ in the right panel of Fig. \ref{1Fig:2}, and it can be seen from this figure that
(a) both modulationally stable and unstable parametric regimes of DIAWs can be obtained; (b) the
modulationally stable parametric regime of the DIAWs increases with an increase in the value of
$q$, and this result is a good agreement with the result of Bains \textit{et al.} \cite{Bains2011};
and (c) the modulationally unstable parametric regime allows to generate first and second order DIARWs.
When $P/Q>0$ and $k_M<k_c$, the MI growth rate $(\Gamma_g)$ is given by \cite{Sultana2011,Chowdhury2018}
\begin{eqnarray}
&&\hspace*{-1.3cm}\Gamma_g=|P|k_M^2\sqrt{\frac{k_c^2}{k_M^2}-1},
\label{1eq:24}
\end{eqnarray}
where $k_M$ is the modulated wave number. Now, we have graphically shown how $\Gamma_g$ varies with
$k_M$ for different values of $\mu_3$  in Fig. \ref{1Fig:3}. It is obvious from Fig. \ref{1Fig:3} that
(a) the growth rate initially increases with $k_M$ and becomes maximum, then reduces to zero;
(b) under the consideration of iso-thermal electrons (i.e., $q=1$ and $\alpha=0$), the maximum value of
the growth rate increases with the number density of the negatively charged dust grains when other
plasma parameters are constant (left panel); (c) $\Gamma_g$ as well as nonlinearity of the plasma system
increases (decreases) with increasing the number density of negative (positive) dust grains (ion) for their
constant charge state under the consideration of non-thermal non-extensive electrons (right panel); (d) the
maximum value of the growth rate increases rigorously in the presence of non-thermal non-extensive electrons
than the presence of iso-thermal electrons.

The existence of the negative ions in the PI plasma rigourously changes the dynamics of the system.
Fig. \ref{1Fig:4} describes the variation of $\Gamma_g$ with $k_M$ for different values of $\mu_3$. It is obvious
from this figure that (a) when we neglect the contribution of negative ions from our plasma system that means only inertial positive
ions, inertialess non-thermal non-extensive electrons, and immobile dust grains are present in the plasma system, then the maximum
value of $\Gamma_g$ decreases with the increase of negative dust grain number density,
and similar effect has also been observed by Javidan and Pakzad \cite{Javidan2014} in a three-component plasma system (left panel);
(b) but when we include negative ions in the plasma system, that means a four-component PI plasma system, then
$\Gamma_g$ increases with increasing the negatively charged dust grains (right panel).
So, the quasi-neutrality condition and associated dynamics of the plasma system are fully changed by negative ions.

The first-order rational solution of Eq. \eqref{1eq:21}, which can predict the concentration of large amount of energy
in a small region of the modulationally unstable parametric regime ($P/Q>0$)
of DIAWs, can be written as \cite{Ankiewicz2009,Guo2012}
\begin{eqnarray}
&&\hspace*{-1.3cm}\Phi_1(\xi,\tau)=\sqrt{\frac{2P}{Q}}\bigg[\frac{4(1+4iP\tau)}{1+16P^2\tau^2+4\xi^2}-1\bigg]\exp(2iP\tau).
\label{1eq:25}\
\end{eqnarray}
We have plotted Eq. \eqref{1eq:25} in Figs. \ref{1Fig:5} and \ref{1Fig:6} to understand the nonlinear properties
of the PI plasma system as well as the mechanism to the formation of DIARWs associated with DIAWs in the modulationally
unstable parametric regime. Figure \ref{1Fig:5} indicates that (a) the amplitude and width of the DIARWs increase
with an increase in the value of $\mu_2$ (both left and right panel); (b) under the consideration of iso-thermal electrons, as
we increase (decrease) the value of negative (positive) ion number density, the amplitude and width of the DIARWs
increase (decrease) when their charge state remain constant (left panel), and this result is similar with the
result of El-Labany \textit{et al.} \cite{El-Labany2012}; (c) similarly, the amplitude and width of the DIARWs
increase with increasing (decreasing)  negative (positive) ion number density when the plasma system has non-extensive non-thermal electrons,
and this result coincides with the result of Abdelwahed \textit{et al.} \cite{Abdelwahed2016};
(d) the direction of the variation of amplitude and width has not been changed with $\mu_2$ under the consideration of iso-thermal or non-thermal non-extensive electrons, but the variation of amplitude and width is severely happened for non-thermal non-extensive electrons than iso-thermal electrons.

It can be seen from the literature that the PI plasma system can support these conditions:
$m_->m_+$ (i.e., $H^+$, $O_2^-$) \cite{Sabry2009,Abdelwahed2016}, $m_-=m_+$ (i.e., $H^+$, $H^-$) \cite{Sabry2009,Abdelwahed2016},
and $m_-<m_+$ (i.e., $Ar^+$, $F^-$) \cite{Sabry2009,Abdelwahed2016}. So, in our present investigation, we have graphically observed the variation
of the electrostatic wave potential with $\mu_1$ in Fig. \ref{1Fig:6}
under the consideration of $m_->m_+$ (i.e., $\mu_1<1$ and left panel) and $m_-<m_+$ (i.e., $\mu_1>1$ and right panel), and it is obvious from this figure that
(a) the shape of the first order DIARWs is so much sensitive to the change of mass and charge state of the positive and negative ions;
(b) the electrostatic wave potential increases with negative ion mass but decreases with the increase of positive ion mass when their charge state
remain constant (both left and right panel); (c) physically, the nonlinearity as well as the amplitude of the DIARWs increases (decreases)
with negative (positive) ion mass, and this result is a good agreement with the result of El-Labany \textit{et al.} \cite{El-Labany2012}.

The second-order rational solution of NLSE can be written as \cite{Ankiewicz2009,Guo2012}
\begin{eqnarray}
&&\hspace*{-1.3cm}\Phi_2(\xi,\tau)=\sqrt{\frac{P}{Q}}\bigg[1+\frac{G_2+iM_2}{T_2}\bigg]\exp(iP\tau),
\label{1eq:26}\
\end{eqnarray}
where
\begin{eqnarray}
&&\hspace*{-1.3cm}G_2(\xi,\tau)=\frac{-\xi^4}{2}-6(P\xi\tau)^2-10(P\tau)^4-\frac{3\xi^2}{2}-9(P\tau)^2+\frac{3}{8},
\nonumber\\
&&\hspace*{-1.3cm}M_2(\xi,\tau)=-P\tau\Big[\xi^4+4(P\xi\tau)^2+4(P\tau)^4-3\xi^2+2(P\tau)^2-\frac{15}{4}\Big],
\nonumber\\
&&\hspace*{-1.3cm}D_2(\xi,\tau)=\frac{\xi^6}{12}+\frac{\xi^4(P\tau)^2}{2}+\xi^2(P\tau)^4+\frac{\xi^4}{8}+\frac{9(P\tau)^4}{2}-\frac{3(P\xi\tau)^2}{2}+\frac{9\xi^2}{16}+\frac{33(P\tau)^2}{8}+\frac{3}{32}.
\nonumber\
\end{eqnarray}
We have graphically shown Eq. \eqref{1eq:26} in  Fig. \ref{1Fig:7}. The nature of the second order DIARWs
is very sensitive to the change of mass and charge state of the PI under the consideration
of $m_->m_+$ (i.e., $\mu_1<1$ and left panel) and $m_-<m_+$ (i.e., $\mu_1>1$ and right panel).
The amplitude of the second order DIARWs associated with DIAWs in the modulationally unstable parametric regime increases (decreases)
with the rise of the mass of the negative (positive) ion for a fixed value of their charge state (both left and right panel).
The physics of this result is that the nonlinearity of the plasma increases with the mass of negative ion but decreases with
the mass of positive ion.

The comparison of the first and second order DIARWs can be observed from Fig. \ref{1Fig:8},
and it is clear from this figure that (a) the second order DIARWs has double structures compared
with the first-order DIARWs; (b) the amplitude  of the second order DIARWs is always greater than the
amplitude of the first order DIARWs; (c) the potential
profile of the second order DIARWs  becomes more spiky (i.e., the taller amplitude and narrower width) than the
first order DIARWs; (d) the second (first) order DIARWs has four (two) zeros symmetrically located
on the $\xi$-axis; (e) the second (first) order DIARWs has three (one) local maxima.
\section{Conclusion}
\label{1sec:Conclusion}
We have studied an unmagnetized realistic PI plasma system having inertialess non-thermal non-extensive
electrons, inertial negative and positive ions, and immobile negatively charged dust grains.
The RPM is used to derive the NLSE, and the nonlinear and dispersive coefficients of the NLSE determine the
stable and unstable parametric regimes of DIAWs. The results that have been found from
our investigation can be summarized as follows:
\begin{itemize}
  \item The angular frequency of the DIAWs decreases (increases) with the increase in the value of non-extensivity (non-thermality)
  of the electrons.
  \item Both modulationally stable and unstable parametric regimes of DIAWs have been observed.
  \item The modulational instability growth rate increases (decreases) with increasing the number density of the negatively charged dust grains
   in the presence (absence) of the negative ions.
  \item The amplitude and width of the DIARWs associated with DIAWs increase (decrease) with the negative (positive) ion mass.
\end{itemize}
It may be noted here that the gravitational and magnetic effects are very important
but beyond the scope of our present work. In future and for
better understanding, someone can investigate the nonlinear
propagation in a four-component PI plasma by considering the
gravitational and magnetic effects.
The results of our present investigation will be useful
in understanding the nonlinear phenomena in both
astrophysical environments such as upper regions of Titan's atmosphere \cite{Massey1976,Sabry2009,Abdelwahed2016,Misra2009,Mushtaq2012},
cometary comae \cite{Chaizy1991}, the ($H^+$, $O_2^-$) and ($H^+$, $H^-$) plasmas in the D and F-regions of Earth's
ionosphere \cite{Massey1976,Sabry2009,Abdelwahed2016,Misra2009,Mushtaq2012},
and also in laboratory plasmas namely, neutral beam sources \cite{Bacal1979}, plasma processing reactors \cite{Gottscho1986},
and Fullerene ($C^+$, $C^-$) \cite{Oohara2003}, etc.
\section*{Acknowledgements}
Authors would like to acknowledge ``UGC research project 2018-2019'' for their financial supports
to complete this work.

\end{document}